\RequirePackage{fix-cm}
\documentclass[twocolumn,epjc3]{svjour3}  
\smartqed  
\RequirePackage{graphicx}
\usepackage{lmodern}
\usepackage{amsmath}
\usepackage{color}

\newcommand{\be}{\begin{equation}}
\newcommand{\ee}{\end{equation}}
\newcommand{\bn}{\begin{eqnarray}}
\newcommand{\en}{\end{eqnarray}}
\newcommand{\bes}{\begin{subequations}}
\newcommand{\ees}{\end{subequations}}

\usepackage{epstopdf}
\usepackage{widetext}
\journalname{Eur. Phys. J. C}
\begin{document}

\title{Evading the non-continuity equation in the $f(R,T)$ formalism}


\author{P.H.R.S. Moraes$^{1,a}$, R.A.C. Correa$^{2,b}$}

\thankstext{1}{e-mail: moraes.phrs@gmail.com}
\thankstext{2}{e-mail: rafael.couceiro@ufabc.edu.br}

\institute{ITA - Instituto Tecnol\'ogico de Aeron\'autica - Departamento de F\'isica, 12228-900, S\~ao Jos\'e dos Campos, S\~ao Paulo, Brazil\\
$^{2}$UNESP, Universidade Estadual Paulista, Campus de Guaratinguet\'{a}, 12516-410, Guaratinguet\'{a}, S\~ao Paulo,
Brazil}

\date{Received: date / Accepted: date}

\maketitle
\begin{abstract}

We present a new approach for the $f(R,T)$ formalism, by exploring the extra terms of its effective energy-momentum tensor $T_{\mu\nu}^{eff}$, namely $\tilde{T}_{\mu\nu}$. Those arise from the consideration of quantum effects, which are usually neglected in general relativity and $f(R)$ theories, and are summed to the usual matter energy-momentum tensor, yielding $T_{\mu\nu}^{eff}=T_{\mu\nu}+\tilde{T}_{\mu\nu}$. Purely from the Bianchi identities, the conservation of both parts of the effective energy-momentum tensor is obtained, rather than the non-conservation of the matter one, originally approached in the $f(R,T)$ theories. In this way, the intriguing scenario of matter-creation, which still lacks observational evidences, is evaded. One is left, then, with two sets of cosmological equations to be solved: the Friedmann-like equations along with the conservation of $T_{\mu\nu}$ and along with the conservation of $\tilde{T}_{\mu\nu}$. We present a physical interpretation for the conservation of the extra terms of the effective energy-momentum tensor, which is related to the presence of stiff matter in the universe. The cosmological features of this approach are presented and discussed as well as the benefits of evading the matter energy-momentum tensor non-conservation. 

\end{abstract}

\section{Introduction}
\label{sec:int}

Alternative cosmological models have been constantly used to solve or at least evade the dark energy (DE) problem \cite{padmanabhan/2008,yoo/2012,wang/2006} found in standard ($\Lambda$CDM) cosmological model. Higher dimensional \cite{overduin/1997,moraes/2016,guo/2015} and $f(R)$ gravity theories \cite{capozziello/2002,starobinsky/2007,lee/2011}, with $f(R)$ indicating an arbitrary function of the Ricci scalar, emerge as optimistic scenarios from which healthy cosmological models can be derived. $f(R)$ gravity in higher dimensions can also generate well behaved models \cite{wu/2014,chakraborty/2015,xu/2015,bazeia/2015,cmsdr/2015}.

In generalized and higher dimensional gravitational models, the extra terms of the field equations\footnote{``Extra terms" when compared to standard gravity field equations.} can induce the effects of a cosmic acceleration, predicted by type Ia supernovae observations \cite{riess/1998,perlmutter/1999} and cosmic microwave background temperature fluctuations \cite{hinshaw/2013}.

A good agreement between theory and cosmological and astrophysical observations can also be obtained from such alternative theories \cite{demianski/2006,lubini/2011,deng/2015b}.

Another reputed alternative gravity model was proposed in \cite{harko/2011}, named the $f(R,T)$ theories of gravity, which present in their field equations extra contributions from both geometry, through a general dependence on $R$, and matter, through a general dependence on $T$, the trace of the energy-momentum tensor (EMT). The $T$-dependence is motivated by the consideration of exotic imperfect fluids or quantum effects.

Cosmological scenarios derived from $f(R,T)$ gravity have been continuously proposed. In \cite{singh/2014}, the authors have introduced bulk viscosity in the $f(R,T)$ formalism within the framework of a flat Friedmann-Robertson-Walker (FRW) model. Since the dependence on $T$ of the gravitational part of the action might come from the consideration of imperfect fluids, the authors have investigated more realistic models, by explicitly taking into account dissipative processes due to viscosity. On this regard, it is known that when neutrinos decoupled in the early universe, matter could have behaved like a viscous fluid.

Moreover, in \cite{houndjo/2014}, a Little Rip model in $f(R,T)$ gravity was investigated. An analysis of $f(R,T)$ models through energy conditions can be found in \cite{sharif/2013}. In \cite{moraes/2016b}, the $f(R,T)$ theories were generalized, by allowing the speed of light to vary, resulting in a primordial scenario alternative to inflation. A complete cosmological scenario was derived from the $f(R,T^{\phi})$ formalism, which is nothing but the $f(R,T)$ theory in the presence of a scalar field, in \cite{ms/2016}. Furthermore, $f(R,T)$ theories of gravity have also been expanded to five dimensions, as explored in \cite{moraes/2014,moraes/2015c,mc/2016,cm/2016}. Very recent contributions of $f(R,T)$ gravity to cosmology, thermodynamics and astrophysics can be found, respectively in \cite{mrc/2016,mmm/2016,mam/2016}.

The well-behaved cosmological models cited above, among others found in the literature, make reasonable and promising to consider the $f(R,T)$ gravity as a possible alternative to General Relativity (GR), from which $\Lambda$CDM cosmological model is derived. Once the gravitational part of the action is generalized, including a general dependence not only on geometrical terms, as in $f(R)$ gravity, but also on material terms through the general dependence on $T$, the new terms of the derived field equations might be responsible for inducting the observed late-time acceleration of the universe expansion with no need of a cosmological constant (CC), which still lacks a convincing physical interpretation \cite{weinberg/1989}. Moreover, as shown in \cite{jamil/2012}, some specific functional forms for $f(R,T)$ may retrieve other cosmological models, as Chaplygin gas and quintessence, manifesting the generic aspect of such a theory of gravity, i.e., different cosmological models found in the literature can be obtained from different particular cases of $f(R,T)$.

$f(R,T)$ theories, as originally proposed, predict a non-conservation of the matter EMT (MEMT), which will be carefully described in Section \ref{sec:frt} below. The non-conservation of the MEMT yields the motion of massive test particles to be non-geodesic, taking place in the presence of an extra-force orthogonal to the four velocity. In \cite{deng/2015}, solar system's bounds on this extra-force were found. In \cite{harko/2014}, the $\nabla^{\mu}T_{\mu\nu}\neq0$ issue was considered to be related to an irreversible matter creation process. It was argued that due to the coupling between matter and geometry predicted in $f(R,T)$ theories, there should be an energy flow between the gravitational field and matter. 

Anyhow, observational evidences of particle creation on a cosmological scale are still missing\footnote{Likewise there are no observational evidences of the predicted extra-force.}. In order to confirm such an intriguing property predicted by $f(R,T)$ gravity (among other theories, such as $f(R,L_m)$ theory \cite{bertolami/2007}, with $L_m$ being the matter lagrangian density), those theories should be tested in a non-usual aspect. If there is creation of matter throughout the universe history, some kind of signature should be imprinted in the cosmic microwave background anisotropy. The classical macroscopic theory predictions in structure formation with linear perturbations may also corroborate or debilitate the scenarios with non-conservation of MEMT.

Such a lack of matter creation observational evidences made some authors to evade the non-continuity equation in $f(R,T)$ gravity. The vanishing of $\nabla^{\mu}T_{\mu\nu}$ was imposed in \cite{chakraborty/2013} and an alternative to the DE problem was obtained. Moreover, the approach has revealed some constraints to the functionality of $f(T)$ in $f(R,T)$. In \cite{sharif/2014} the authors have reconstructed $f(R,T)$ gravity for a specific model that permitted the standard continuity equation to hold. The dynamics and stability of $f(R,T)$ theory for de-Sitter and power-law expansions of the universe with conservation of MEMT were discussed in \cite{baffou/2014}.

In the present approach we will also evade the non-conservation of MEMT predicted in $f(R,T)$ theories. However there will be no imposition of conservation. The non-vanishing of the MEMT covariant derivative will be evaded purely from mathematical identities of GR. 

The article is organized as follows. In Section \ref{sec:frt} a brief review of $f(R,T)$ gravity is given. The equation for the non-conservation of the MEMT is indicated. In Section \ref{sec:c2} we distinguish the presence of extra terms in the effective energy-momentum tensor (EEMT) in $f(R,T)$ formalism and construct a scenario in which there is no violation of the continuity equation. General cosmological solutions to this new approach are presented in Section \ref{sec:ns}. In Section \ref{sec:asm} we physically interpret the extra terms of the EEMT. Section \ref{sec:peemt} is devoted to discuss the benefits of having the non-conservation of the MEMT evaded in a given theory. Discussion and conclusions are presented in Section \ref{sec:dc}.

\section{A brief review of the $f(R,T)$ gravity}
\label{sec:frt}

Recently proposed by T. Harko and collaborators \cite{harko/2011}, the $f(R,T)$ theory of gravity assumes the gravitational part of the action depends on an arbitrary function of $R$ and $T$. According to the authors, the dependence on $T$ may be induced by the consideration of exotic fluids or quantum effects. The total action in such a theory is given by

\begin{equation}\label{eqn:frt1}
S=\frac{1}{16\pi}\int d^{4}xf(R,T)\sqrt{-g}+\int d^{4}xL_m\sqrt{-g}.
\end{equation}
In (\ref{eqn:frt1}), $f(R,T)$ is the arbitrary function of $R$ and $T$, $g$ is the metric determinant and $L_m$ is the matter lagrangian density. Moreover, we will assume $c=G=1$.

The MEMT is written as

\begin{equation}\label{eqn:frt2}
T_{\mu\nu}=g_{\mu\nu}L_m-2\frac{\partial L_m}{\partial g^{\mu\nu}}.
\end{equation}
We will assume the matter source is a perfect fluid (PF) in (\ref{eqn:frt2}) and the universe is homogeneous and isotropic at cosmological scales, i.e., its geometry is described by an FRW metric. We will also consider a flat universe, in accordance with recent cosmic microwave background observations \cite{hinshaw/2013}, and work with the case $f(R,T)=R+2\lambda T$, with $\lambda$ being a constant. Originally suggested by the $f(R,T)$ gravity authors \cite{harko/2011}, such a functional form for $f(R,T)$ has been extensively used to obtain $f(R,T)$ cosmological solutions (check \cite{singh/2014,moraes/2016b,moraes/2014,moraes/2015c,mc/2016} among many others). Moreover, by assuming $f(R,T)=R+2\lambda T$, such an $f(R,T)$ functional form benefits from the fact that one can recover GR just by letting $\lambda$ to be null.

Here it is worth stressing that, as in the present case, when the functionality $f(R,T)=R+f(T)$, with $f(T)$ being a function of $T$, is taken into account, the derived models do not predict a matter-geometry coupling, although this case is sometimes considered as a ``minimal coupling" in the literature. A product between $R$ and $T$ or functions of them in $f(R,T)$ would, rather, lead to an explicit (non-minimal) coupling. Such a hitherto non-investigated issue of $f(R,T)$ theory of gravity will be important to this article purposes and concluding remarks, and shall be appreciated later on.

The assumptions above yield, for the variation of the action (\ref{eqn:frt1}) with respect to the metric, the following field equations 

\begin{equation}\label{eqn:frt3}
G_{\mu\nu}=8\pi T_{\mu\nu}+2\lambda(T_{\mu\nu}+pg_{\mu\nu})+\lambda Tg_{\mu\nu},
\end{equation}
for which $G_{\mu\nu}$ is the usual Einstein tensor and $p$ is the pressure of the universe.

A notorious feature about $f(R,T)$ theory is the non-nullity of the MEMT covariance divergence. Recently, the authors in \cite{alvarenga/2013,barrientos/2014} have recalculated $\nabla^{\mu}T_{\mu\nu}$, by arguing that T. Harko et al. \cite{harko/2011} missed an essential term which has consequences in the equation of motion of test particles. From such an argumentation, the corrected relation for $\nabla^{\mu}T_{\mu\nu}$ when $f(R,T)=R+2\lambda T$ is

\begin{equation}\label{eqn:frt4}
\nabla^{\mu}T_{\mu\nu}=\frac{2\lambda}{2\lambda-8\pi}\left[\nabla^{\mu}(2T_{\mu\nu}+pg_{\mu\nu})+\frac{1}{2}g_{\mu\nu}\nabla^{\mu}T\right].
\end{equation}
Note that, as required, the case $\lambda=0$ retrieves GR in both (\ref{eqn:frt3}) and (\ref{eqn:frt4}).

\section{The conservation of the effective energy-momentum tensor in $f(R,T)$ gravity}
\label{sec:c2}

The {\it lhs} of the field equations (\ref{eqn:frt3}) is exactly the same as in GR, i.e., given by the Einstein tensor. The {\it rhs} clearly presents some extra terms when compared to standard gravity. In order to obtain an accelerated expanding universe in standard cosmology, one has to assume that $\sim70\%$ of the universe composition is in the form of some exotic fluid dubbed DE, which may enter the GR field equations in the form of a CC EMT. The extra terms in (\ref{eqn:frt3}) may, in a sense, play the role of the CC, however, evading the DE problem quoted above. 

As carefully highlighted in the previous section, Eq.(\ref{eqn:frt2}) refers to MEMT. By keeping that in mind we can rewrite (\ref{eqn:frt3}) as

\begin{equation}\label{eqn:hemt1}
G_{\mu\nu}=8\pi T_{\mu\nu}^{eff},
\end{equation}
with $T_{\mu\nu}^{eff}=T_{\mu\nu}+\tilde{T}_{\mu\nu}$ and 

\begin{equation}\label{eqn:hemt2}
\tilde{T}_{\mu\nu}\equiv\frac{1}{8\pi}[2\lambda(T_{\mu\nu}+pg_{\mu\nu})+\lambda Tg_{\mu\nu}]
\end{equation}
being the extra terms of the EEMT $T_{\mu\nu}^{eff}$.

By considering a PF in Equation (\ref{eqn:frt2}) yields $T_{\mu\nu}=diag(\rho,-p,-p,-p)$, with $\rho$ being the matter-energy density of the universe and whose trace is given by $T=\rho-3p$. Yet, from (\ref{eqn:hemt2}), 

\begin{equation}\label{eqn:hemt3}
\tilde{T}_{00}=\frac{\lambda}{8\pi}(3\rho-p),
\end{equation}
\begin{equation}\label{eqn:hemt4}
\tilde{T}_{11}=-\frac{\lambda}{8\pi}(\rho+p),
\end{equation}
\begin{equation}\label{eqn:hemt5}
\tilde{T}_{22}=\tilde{T}_{33}=\tilde{T}_{11},
\end{equation}
and $\tilde{T}=-(\lambda/2\pi)p$. In this way, the fluid described by Eq.(\ref{eqn:hemt2}) permeates the universe along with the ordinary PF which is often considered as the (only) matter source of the universe. Effectively, there is one fluid permeating the universe, whose density is given by the sum of the densities of both fluids determined above. We shall revisit this question later on.

The non-null components of (\ref{eqn:hemt1}) yield the Friedmann-like equations

\begin{equation}\label{eqn:hemt6}
3\left(\frac{\dot{a}}{a}\right)^{2}=(8\pi+3\lambda)\rho-\lambda p,
\end{equation}
\begin{equation}\label{eqn:hemt7}
2\frac{\ddot{a}}{a}+\left(\frac{\dot{a}}{a}\right)^{2}=\lambda\rho-(8\pi+\lambda)p,
\end{equation}
with dots representing derivatives with respect to time. As required, when $\lambda=0$ the standard Friedmann equations of $\Lambda$CDM cosmology are retrieved.

When one breaks the EEMT in two parts, one responsible for the matter in the universe while the other comes from the dependence of the action on $T$, a remarkable issue about the MEMT non-conservation, predicted by $f(R,T)$ theories, arises: note that what makes $\nabla^{\mu}T_{\mu\nu}\neq0$ in (\ref{eqn:frt4}) is the indistinct presence of $\tilde{T}_{\mu\nu}$ defined in Eq.(\ref{eqn:hemt2}). However, the application of the Bianchi identities ($\nabla^{\mu}G_{\mu\nu}=0$) in (\ref{eqn:hemt1}) yields $\nabla^{\mu}[8\pi(T_{\mu\nu}+\tilde{T}_{\mu\nu})]=0$, or:

\begin{equation}\label{eqn:hemt8}
\nabla^{\mu}T_{\mu\nu}=0
\end{equation}
and

\begin{equation}\label{eqn:hemt9}
\nabla^{\mu}\tilde{T}_{\mu\nu}=0.
\end{equation}
Therefore one has two sets of three equations with three unknowns to be solved: (\ref{eqn:hemt6}), (\ref{eqn:hemt7}), (\ref{eqn:hemt8}) and (\ref{eqn:hemt6}), (\ref{eqn:hemt7}), (\ref{eqn:hemt9}). 

Note that by approaching the $f(R,T)$ gravity from such a perspective indeed evades the non-continuity equation originally predicted in the formalism. 

\section{Cosmological solutions from the effective energy-momentum tensor conservation}
\label{sec:ns}

The evaluation of (\ref{eqn:hemt8}) yields the well-known continuity equation of cosmology:

\begin{equation}\label{eqn:hemt10}
\dot{\rho}+3\frac{\dot{a}}{a}(\rho+p)=0.
\end{equation}

By recombining Eqs.(\ref{eqn:hemt6}), (\ref{eqn:hemt7}) and (\ref{eqn:hemt10}), it is straightforward to obtain the following differential equation for the scale factor

\begin{eqnarray}\label{hemt11}
&&L(\lambda)\left[\frac{\ddot{a}}{a}+\frac{1}{2}\left(\frac{\dot{a}}{a}\right)^{2}\right]+\lambda\left[\frac{8\pi}{3}\left(\frac{\dddot{a}}{\dot{a}}-\frac{\dot{a}}{a}\right)-\lambda\frac{\ddot{a}}{a}\right]+\nonumber \\
&&(8\pi+4\lambda)\left[(8\pi+3\lambda)\frac{\ddot{a}}{a}+4\pi\left(\frac{\dot{a}}{a}\right)^{2}\right]=0,
\end{eqnarray}
with $L(\lambda)\equiv\lambda^{2}-(8\pi+3\lambda)(8\pi+\lambda)$.

By developing (\ref{eqn:hemt9}) yields

\begin{equation}\label{eqn:hemt12}
\dot{\rho}=\dot{p}.
\end{equation}

Eq.(\ref{eqn:hemt12}), if integrated, reveals the presence of stiff matter (SM) in the universe. Such a prediction of the present model shall be revisited later on.

By recombining Eqs.(\ref{eqn:hemt6}), (\ref{eqn:hemt7}) and (\ref{eqn:hemt12}), one obtains

\begin{equation}\label{eqn:hemt13}
\frac{8\pi}{3}\left(\frac{\dddot{a}}{\dot{a}}-\frac{\dot{a}}{a}\right)-\lambda\frac{\ddot{a}}{a}=\frac{1}{3}\frac{L(\lambda)}{8\pi-2\lambda}\left[\frac{\ddot{a}}{a}-\left(\frac{\dot{a}}{a}\right)^{2}\right].
\end{equation}

Substituting Equation (\ref{eqn:hemt13}) into (\ref{hemt11}), we have
\begin{equation}
\left(  \frac{\ddot{a}}{a}\right)  +\frac{\Lambda_{2}}{\Lambda_{1}}\left(
\frac{\dot{a}}{a}\right)  ^{2}=0,\label{as2}
\end{equation}
where we are using the definitions

\begin{align}
\Lambda_{1}  & \equiv L(\lambda)\left(  1+\frac{\lambda}{24\pi+6\lambda}\right)
+(8\pi+3\lambda)(8\pi+4\lambda),\label{as3}\\
\Lambda_{2}  & \equiv L(\lambda)\left(  \frac{1}{2}+\frac{\lambda}{24\pi+6\lambda
}\right)  +4\pi(8\pi+4\lambda).\label{as4}
\end{align}

In this way, after straightforward manipulations, we can put Eq.(\ref{as2}) in the form
\begin{equation}
\frac{d}{dt}\left(  \ln\dot{a}+\varsigma\ln a\right)  =0,\label{as5}
\end{equation}
with $\varsigma\equiv\Lambda_{2}/\Lambda_{1}$. Hence, by solving the
above equation and applying some manipulations, we conclude that the scale
factor can be written as
\begin{equation}
a(t)=a_{0}t^{\mathcal{G}},\label{as6}
\end{equation}
where $a_{0}$ is an arbitrary constant of integration and $\mathcal{G}\equiv1/(1+\varsigma)$. 

\begin{figure}[ht!]
\vspace{0.3cm}
\centering
\includegraphics[height=5cm,angle=00]{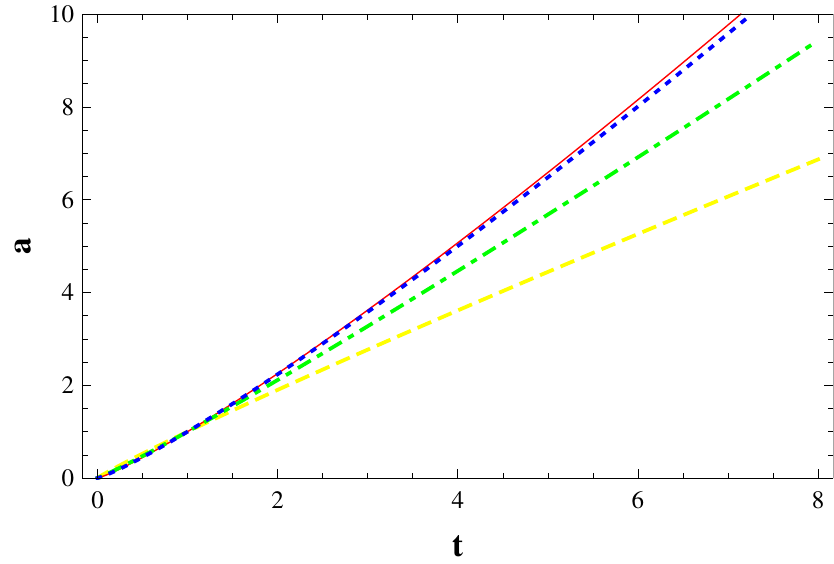}
\caption{Time evolution of the scale factor presented in Eq.(\ref{as6}). The (blue) dotted line stands for $\lambda=\pi$, while the (green) dot-dashed, (red) solid and (yellow) dashed curves represent $\lambda=-\pi$, $\lambda=5\pi$ and $\lambda=-5\pi$, respectively. In all curves we are taking $a_0=1$.}
\label{fig1}
\end{figure}

The evolution of the scale factor (\ref{as6}) in time is depicted in Fig.\ref{fig1} below for different values of $\lambda$.

From Eq.(\ref{as6}) we can calculate the Hubble parameter $H=\dot{a}/a$ and the deceleration parameter $q=-\ddot{a}a/\dot{a}^{2}$.

\begin{figure}[]
\vspace{0.3cm}
\centering
\includegraphics[height=5cm,angle=00]{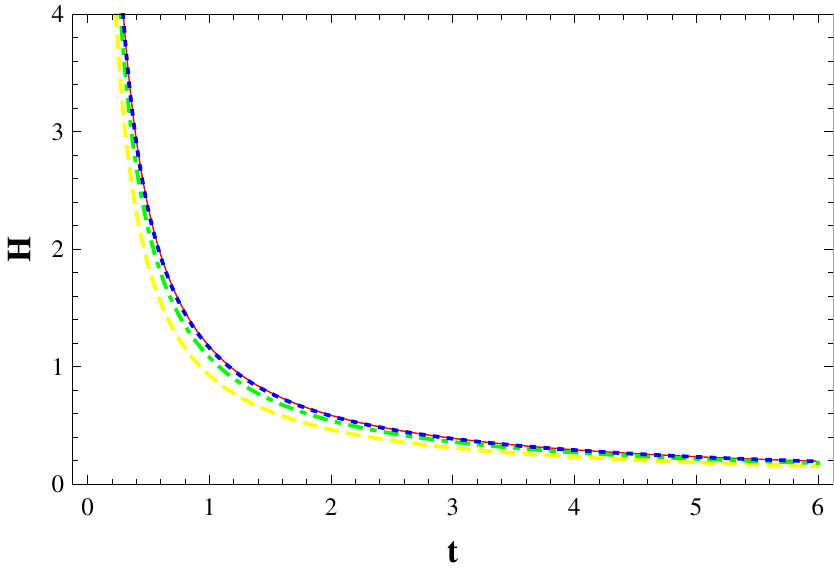}
\caption{Time evolution of the Hubble parameter $H=\dot{a}/a$, with $a$ given by Eq.(\ref{as6}). The (blue) dotted line stands for $\lambda=\pi$, while the (green) dot-dashed, (red) solid and (yellow) dashed curves represent $\lambda=-\pi$, $\lambda=5\pi$ and $\lambda=-5\pi$, respectively.}
\label{fig2}
\end{figure}

The Hubble parameter time evolution is depicted in Fig.\ref{fig2} above for different values of $\lambda$.

Moreover, the present solutions predict a cosmic acceleration of the universe ($q<0$) in the following ranges:

\begin{equation}\label{hd1}
\lambda<-\frac{\pi}{2}\left(  7+\sqrt{17}\right),
\end{equation}
\begin{equation}\label{hd2}
\lambda>-\frac{\pi}{2}\left(  7-\sqrt{17}\right),
\end{equation}
\begin{equation}\label{hd3}
-\frac{8\pi}{29}\left(  11+\sqrt{5}\right)  <\lambda<-\frac{8\pi}{29}\left(  11-\sqrt{5}\right).
\end{equation}

In possession of Eq.(\ref{as6}) we can also obtain solutions for the energy density and pressure of the universe.

We can solve the Friedmann-like equations of the model for $\rho$ by using Eqs.(\ref{eqn:hemt10}) and (\ref{eqn:hemt12}), obtaining, in this way, solutions for the usual PF and for the SM fluid (SMF), respectively. Effectively, we have one fluid whose matter-energy density is given by the sum of those (check Fig.\ref{fig3} below).

\begin{figure}[ht!]
\vspace{0.3cm}
\centering
\includegraphics[height=5cm,angle=00]{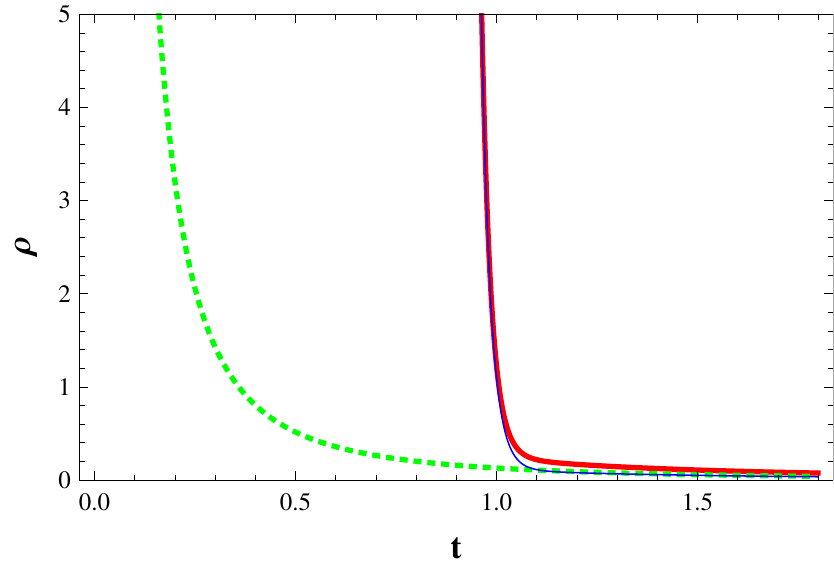}
\caption{Time evolution of the matter-energy density of the universe for $\lambda=\pi$. The (green) dotted curve stands for the SMF contribution. The (blue) thin solid line represents the ordinary PF contribution and the (red) thick solid line is the effective energy density of the universe, given by the sum of the contributions.}
\label{fig3}
\end{figure} 

By substituting the above result in the Friedmann-like equations and rewriting Eqs.(\ref{eqn:hemt10}) and (\ref{eqn:hemt12}) for $p$, we obtain the solutions for the pressure of the universe that can be seen in Fig.\ref{fig4} below.

\begin{figure}[!]
\vspace{0.3cm}
\centering
\includegraphics[height=5cm,angle=00]{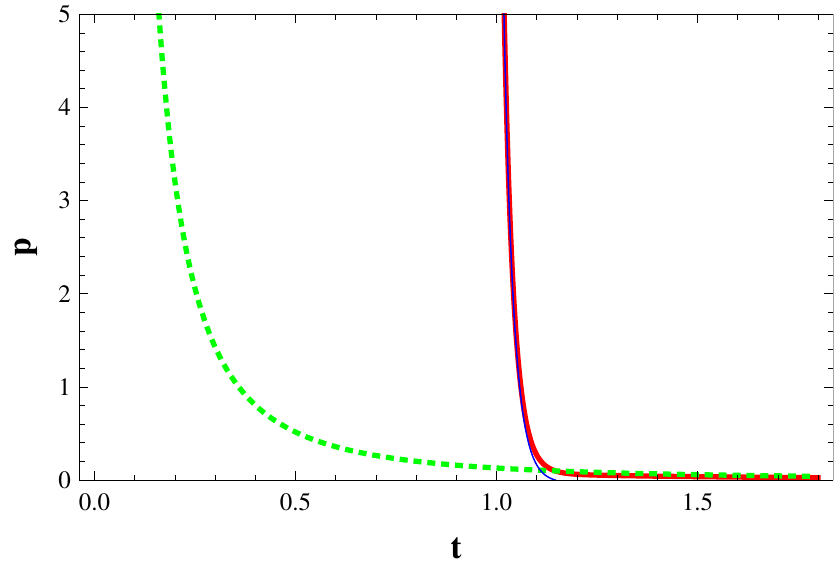}
\caption{Time evolution of the pressure of the universe for $\lambda=\pi$. As in Fig.3, the (green) dotted curve stands for the SMF contribution. The (blue) thin solid line represents the ordinary PF contribution and the (red) thick solid line is the effective pressure of the universe, given by the sum of the contributions.}
\label{fig4}
\end{figure} 

\section{The arising of stiff matter in the universe}\label{sec:asm}

Eq.(\ref{eqn:hemt12}), after integrated, describes a SMF. It is not the first time that $f(R,T)$ gravity reveals the possibility of existence of such kind of matter in the universe. In \cite{moraes/2015c}, by making one of the integration constants of the cosmological solutions to vanish, one obtains an SM dominated scenario. In \cite{farasat_shamir/2015,mahanta/2014} the presence of SMF permeating the universe was also predicted in the $f(R,T)$ scenario.

Some early universe models indicate that there may have existed a phase prior to the radiation-dominated era in which our universe dynamics was dominated by a gas of baryons with equation of state (EoS) $p=\rho$ which interacted through a vector-meson field \cite{rama/2009}. 

The conjecture of a primordial SM era first appeared in \cite{zeldovich/1972}. The presence of SM in cosmological scenarios may explain the baryon asymmetry and the density perturbations of the right amplitude for the large scale structure formation in the universe, as shown in \cite{joyce/1998}. It can play an important role in the spectrum of inflationary gravitational waves \cite{sahni/2001} and characterize the EoS of neutron stars \cite{suleimanov/2011}. A thermodynamics analysis of an FRW universe with bulk viscous SMF was made in \cite{mathew/2014}. An article presenting a broad study of cosmology with an SM era was recently written \cite{chavanis/2015}. SMF has also appeared in anisotropic cosmological models \cite{mak/2005,chen/2001,rama/2009b}. Moreover, SM occurs in the relativistic scalar fields approach when their kinetic energy dominates the potential energy. A primordial SM era is fundamental for any model based on a relativistic scalar field \cite{li/2014}.

Our purpose in Section \ref{sec:c2} was to propose a form of evading the EMT non-conservation issue surrounding $f(R,T)$ gravity theory. By doing this, Eqs.(\ref{eqn:hemt8})-(\ref{eqn:hemt9}) have arose. By developing them, we obtained Eqs.(\ref{eqn:hemt10}) and (\ref{eqn:hemt12}). Eq.(\ref{eqn:hemt10}) stands for the usual conservation of matter represented by a PF. On the other hand, Eq.(\ref{eqn:hemt12}) predicts the existence of SM in the universe.

As in another SM cosmological models (check \cite{chavanis/2015}, for instance), everything happens as if the universe were composed of two non-interacting fluids, in our case, one respecting Eq.(\ref{eqn:hemt10}) and another submissive to Eq.(\ref{eqn:hemt12}). Effectively, we have just one fluid whose energy density is given by the sum of the solutions of Eqs.(\ref{eqn:hemt10}) and (\ref{eqn:hemt12}) (check Fig.\ref{fig3} above). 

\section{Evading the non-continuity equation}
\label{sec:peemt}

Despite the lack of observational evidences corroborating scenarios with continuum or episodic creation of particles in cosmological scales, some efforts on this subject have been made \cite{singh/2016,quintin/2014}.

One might wonder how the evasion of non-conservation of the MEMT which is described in Section \ref{sec:c2} can coexist with $f(R,T)$ models whose MEMT is not conserved. We argue about this issue below.

A matter-curvature coupling is the mechanism considered to be responsible for gravitationally induce particle production in $f(R,T)$ or $f(R,L_m)$ theories \cite{harko/2014,bertolami/2007,harko/2015}. The coupling of matter with higher order derivative curvature terms can be interpreted as an exchange of energy and momentum between them, which induces a gravitational particle production. In other words, the matter-curvature coupling generates an irreversible flow from the gravitational field to created matter constituents and the second law of thermodynamics requires that geometric curvature transforms into matter (check \cite{harko/2015}).

In $f(R,L_m)$ theory, the functional form $f_1(R)+[1+\lambda f_2(R)]L_m$, with $f_1(R)$ and $f_2(R)$ being arbitrary functions of $R$, was considered in the study of gravitationally induced particle creation \cite{harko/2015}. Recalling the argumentations above, it is clear that such a model indeed predicts a coupling of matter and geometry, through the product $f_2(R)L_m$, justifying, at a theoretical level, a scenario with matter creation. Such a coupling shall be prevented when $\lambda=0$.

Departing from the case in Ref.\cite{harko/2015}, the functional form used in the present article, $f(R,T)=R+2\lambda T$, does not predict a matter-geometry coupling (recall that there is no product between material terms, proportional to $T$, and geometrical terms, proportional to $R$), thus there is no mechanism able to explain geometric curvature transforming into matter.

Therefore when there is no coupling between matter and geometry, as in the $f(R,T)=R+2\lambda T$ theory, one is capable of escaping the argument of creation of matter in the universe through the approach above, which is worth since so far there is no observational evidences of such a phenomenon.

\section{Discussion and conclusions}
\label{sec:dc}

There is plenty of motivations for the search of alternative cosmological models nowadays. That happens because, although $\Lambda$CDM cosmological model is able to provide a good agreement between theoretical predictions and observations, it is surrounded by shortcomings and drawbacks (check \cite{padmanabhan/2008,yoo/2012,wang/2006}). 

In this article we have approached the $f(R,T)$ theory of gravity from a cosmological perspective, proposing a form of surpassing the non-conservation of MEMT, which is originally predicted in such an alternative theory. 

Our proposal was to establish an alternative form of treating the EEMT of $f(R,T)=R+2\lambda T$ gravity, by distinguishing the presence of a fluid described by Eq.(\ref{eqn:hemt2}). Such an establishment implies the continuity equations (\ref{eqn:hemt8})-(\ref{eqn:hemt9}).

The integration of Eq.(\ref{eqn:hemt12}) has revealed the presence of SMF permeating the universe along with the usual PF described by Eq.(\ref{eqn:hemt10}). Note that two-fluid cosmological models have been proposed for a long time in the literature \cite{dunn/1989,coley/1986,coley/1992}. For recent references on this regard, check \cite{harko/2011b,lip/2011,aoki/2014}. Even three-fluid cosmological models have already been constructed \cite{tsamparlis/2012}.

Recall that as argued by the $f(R,T)$ gravity authors, the motivation to insert a function of $T$ in the gravitational part of the action lies on the consideration of exotic or imperfect fluids which are usually neglected in GR or $f(R)$ gravity. In this way, the rising of SMF predicted by the integration of Eq.(\ref{eqn:hemt12}) is not surprising. Moreover, recall that SMF existence was already predicted by $f(R,T)$ models in \cite{moraes/2015c,farasat_shamir/2015,mahanta/2014} and also in \cite{singh/2014b}.

From our approach, besides predicting the existence of SMF, we could evade the MEMT non-conservation in $f(R,T)$ gravity. The main motivation to such an evasion is the lack of observational evidences of particle creation in cosmological scales.

One might wonder if the MEMT non-conservation evasion of the present approach conflicts with the original predictions of $f(R,T)$ gravity. The answer is no. In this article we have shown that when there is no product between $f(T)$ and $R$, namely when there is no matter-geometry coupling, one is able to break the EEMT of $f(R,T)$ cosmology in two terms. Then, by applying the Bianchi identities only, we have shown that both conserve.

It is worth quoting here that in $f(R,T)$ gravity, departing from $f(R,L_m)$ models, there is no product between $f(R)$ (or simply $R$, as in our case) and $L_m$. This can be checked by recalling Eq.(\ref{eqn:frt1}), where the sum of the matter action (proportional to $L_m$) to the gravitational action (proportional to $f(R,T)$) is explicit. 

Cosmologically, the present model has shown to be very healthy and well-behaved. Our cosmological solutions have been constructed from the scale factor of Eq.(\ref{as6}), which was obtained from the method presented in Section \ref{sec:ns}. Fig.\ref{fig1} shows the time evolution of the scale factor for different values of $\lambda$. We can see that the (yellow) dashed curve, with $\lambda=-5\pi$ departs significantly from the other curves. The implications of that feature are discussed some paragraphs below.

Fig.\ref{fig2} shows the time evolution of the Hubble parameter $H=\dot{a}/a$ for different values of $\lambda$. The curves plotted for the Hubble parameter are cosmologically healthy for the following reasons: firstly, independently of the value of $\lambda$, the Hubble parameter always assume only positive values, which is in agreement with an expanding universe; secondly, in standard cosmology it is well known that the Hubble parameter evolves as $1/t_H$, with $t_H$ being the Hubble time. Such a standard behaviour is being recovered in the present approach, as one can see in Fig.\ref{fig2} above.

The present model is also able to predict the current accelerated expansion of the universe since it predicts negative values of the deceleration parameter $q$ for most of the values considered by us for $\lambda$ ($\lambda=\pi$, $\lambda=-\pi$, $\lambda=5\pi$). In fact, in order to obtain $q<0$, $\lambda$ just needs to obey one of the constrains (\ref{hd1})-(\ref{hd3}), and $\lambda=-5\pi$, which represents the departing curve of the scale factor in Fig.\ref{fig1}, does not respect any of those. 

From Eq.(\ref{as6}), we were also able to plot the matter-energy density (Fig.\ref{fig3}) and pressure of the universe (Fig.\ref{fig4}) according to the model.

Fig.\ref{fig3} depicts the time evolution of the SM density, represented by a (green) dotted-curve, of the ordinary PF density, represented by a (blue) thin solid line, and of the effective density, represented by a (red) thick solid line. It predicts the existence of SMF prior to the usual PF. As quoted in Section \ref{sec:asm}, some early universe models  indicate that there may have existed a phase prior to the radiation-dominated era in which our universe dynamics was dominated by a fluid with EoS $p=\rho$. Such an indication is corroborated by Fig.\ref{fig3}.

SMF density tends to $0$ as time passes by. The low, but non-null, SM density contribution for high values of time may be related with the presence of neutron stars and even other compact astrophysical objects in the universe, which can have SM in their core (check \cite{suleimanov/2011} and also \cite{drago/2014}).

To finish, we would like to point that in Fig.\ref{fig4}, the pressure of the ordinary PF assumes a null value long before the SM and effective pressures do (note that the (blue) thin and (red) thick lines tends asymptotically to $0$). On this regard, it is worth mentioning that after the radiation-dominated era, the dynamics of the universe became dominated by dark matter\footnote{We can say that at that time, the universe dynamics became dominated by {\it matter} instead of {\it dark matter}, in which {\it matter} would stand for the contribution from both dark matter and baryonic matter. However, the baryonic matter contribution can be neglected when compared to the dark matter portion predicted to exist in the universe (check \cite{hinshaw/2013}).} (DM) \cite{farooq/2013,nojiri/2006,ms/2014}. 

The nature of DM is still unknown, but in $\Lambda$CDM cosmological model, it is modeled as a pressureless fluid, i.e., a fluid  with $p=0$. Such a modeling is appropriate when considering DM to be made of weakly interacting massive particles (usually known as WIMPs) with mass in the GeV-TeV range \cite{chavanis/2015}. 

In this way the DM era is being appropriately predicted in Fig.\ref{fig4}, since we are living in an epoch in which the PF permeating the universe has $p=0$.

In $\Lambda$CDM model, the present EoS of the PF reads $p\sim-\rho$ \cite{hinshaw/2013}. This happens because in order to theoretically predict the cosmic acceleration in standard cosmology, one invokes the CC, which is physically interpreted as an exotic fluid with negative pressure (which would cause the acceleration of the universe expansion). On the other hand, we did not invoke the CC in the present model. Since anyhow we have obtained an accelerated expansion of the universe ($q<0$), we can infer that in this model such a phenomenon occurs as a consequence of the consideration of terms proportional to $T$ in the gravitational part of the action and consequently in the field equations. In other words, not only the presence of extra geometrical terms in the gravitational part of the action can account for the cosmic acceleration, as shown in several $f(R)$ models, such as \cite{vacaru/2014,hu/2007,nojiri/2011,amendola/2007,navarro/2007,song/2007,capozziello/2005,nojiri/2013,bamba/2010}, but also the consideration of extra material terms.

\begin{acknowledgements}
PHRSM would like to thank S\~ao Paulo Research Foundation (FAPESP), grant 2015/08476-0, for financial support. PHRSM would also like to thank Drs. D. Momeni, H. Ludwig and J.D.V. Arba\~nil for some helpful discussions about matter-geometry coupling, particle creation and imperfect fluids. RACC thanks CAPES for financial support. The authors also thank the anonymous referee for his/her constructive criticism to our paper. His/her suggestions have significantly improved the physical content of this work.
\end{acknowledgements}



\end{document}